\documentclass[epj]{svjour}
\usepackage{lineno}
\usepackage{color}

\usepackage{graphicx}
\usepackage{amssymb}
\usepackage{epsfig}

\newcommand{\jpsi}{{\rm J}/\psi}

\newcommand{\nin}{{N^{\rm in}}}
\newcommand{\nout}{{N^{\rm out}}}
\newcommand{\sigmain}{{\sigma^{\rm in}}}
\newcommand{\sigmaout}{{\sigma^{\rm out}}}
\begin{document}
\title{First Measurement of $J/\psi$ Azimuthal Anisotropy in PHENIX at Forward Rapidity in Au+Au Collisions at $\sqrt{s_{\rm NN}}=200$~GeV}
\author{Catherine Silvestre\inst{1}
\thanks{\emph{\email{silvestr@rcf.rhic.bnl.gov}}} (for the PHENIX Collaboration)%
}                     
\offprints{}          
\institute{IRFU, CEA-Saclay\\F-91191, Gif-sur-Yvette, France}
\date{Received: date / Revised version: date}
%
\abstract{
The PHENIX experiment has shown that $\jpsi$s are suppressed in central Au+Au collisions at a center of mass energy per nucleon-nucleon collision $\sqrt{s_{NN}}=200$~GeV, and that the suppression is larger at forward than at mid-rapidity. Part of this difference may be explained by cold nuclear matter effects but the most central collisions suggest that regeneration mechanisms could be at play. In 2007, PHENIX collected almost four times more Au+Au collisions at this energy than used for previous published results. Moreover, the addition of a new reaction plane detector allows a much better analysis of the $\jpsi$ behavior in the azimuthal plane. Since a large elliptic flow has been measured for open charm, measuring $\jpsi$ azimuthal anisotropies may give a hint if $\jpsi$ are recombined in the expanding matter. First PHENIX results of $\jpsi$ elliptic flow as a function of transverse momentum at forward rapidity are presented in this article. The analysis is detailed and results are compared to mid-rapidity PHENIX preliminary results as well as to predictions.
%
} 
\authorrunning{Catherine Silvestre}
\titlerunning{$\jpsi$ $v_2$ at forward rapidity in Au+Au collisions by PHENIX}
\maketitle
\section{Introduction}
\label{intro}
A non-central collision creates a spatial anisotropy of the overlapping interaction region. In a thermodynamical picture, the asymmetric distribution of the initial energy density causes a larger pressure gradient in the shortest direction of the ellipsoidal medium which can lead to an anisotropic azimuthal emission of particles. Such an \emph{elliptic flow} is quantified by measuring the second Fourier coefficient, $v_2$, of the produced particles in the $\phi$ direction, with respect to the reaction plane angle, $\psi$~\cite{Ollitrault:1997di,PhysRevC.58.1671}: 
\begin{equation}
\frac{N}{d(\phi-\psi)}=A\cdot[1+ 2\cdot v_2\cdot cos(2(\phi-\psi))]
\end{equation} 
A positive elliptic flow has been measured for light mesons~\cite{Afanasiev:2007tv}, which supports a picture with a rapid thermalization of the system.

The $c\bar c$ pair that eventually forms a $\jpsi$ is produced in the early part of the collision (through gluon fusion at RHIC energies). These are called direct $\jpsi$s. Part of the $\jpsi$ yield also comes from feed down of excited states ($\psi'$ and $\chi_c$) and can contribute to up to $\sim 40\%$~\cite{Morino:2008nc}. In heavy ion collisions, charmonia production rate can be reduced, for example, by gluon shadowing due to nuclear parton distribution modification~\cite{Eskola:2001gt,deFlorian:2003qf}, nuclear absorption~\cite{Arleo:2006qk} or color screening in a quark gluon plasma~\cite{MatsuiSatz}. In addition, interactions with comovers~\cite{Tywoniuk:2008ie} might also dissociate $\jpsi$s. On the other hand, given the large density of uncorrelated charm quarks in the hot medium at RHIC, these may recombine and enhance the $\jpsi$ yield~\cite{thews,Grandchamp:2004tn}. 

\begin{figure}[h]
\centering\resizebox{0.45\textwidth}{!}{%
  \includegraphics{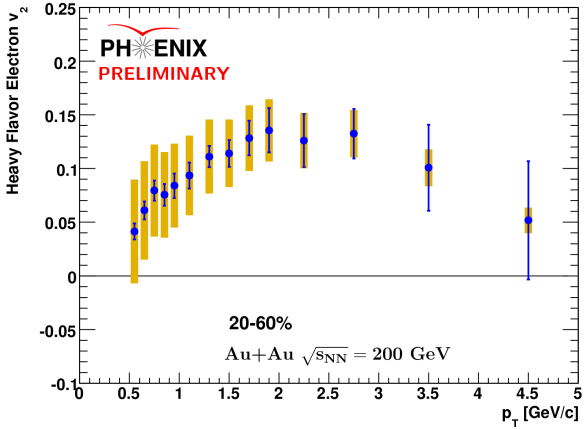}
}
\caption{\label{fig:open_elec_v2}Non-photonic electrons $v_2$ as a function of $p_T$ for the centrality 20-60\% at mid-rapidity ($|y|<0.35$)}      
\end{figure}

A way to experimentally disentangle direct from regenerated $\jpsi$s is to look for $\jpsi$ elliptic flow. Indeed PHENIX has measured a large elliptic flow for open charm from non-photonic electrons at mid-rapidity~\cite{Dion} as shown in Figure~\ref{fig:open_elec_v2} for the same centrality class used for the $\jpsi$ $v_2$ reported here. If regeneration is strong in this rapidity region, one would expect $\jpsi$s to inherit the positive elliptic flow of the open charm from which they are formed. Assuming similar open charm elliptic flow at forward rapidity, measuring $\jpsi$ $v_2$ at various rapidities could indicate if the difference seen in their nuclear modification factor comes from regeneration. If no $\jpsi$ elliptic flow is measured, there is little chance that the measured $\jpsi$s come from regeneration.

\section{Detectors}
\begin{figure}[h]
\centering\resizebox{0.4\textwidth}{!}{%
  \includegraphics{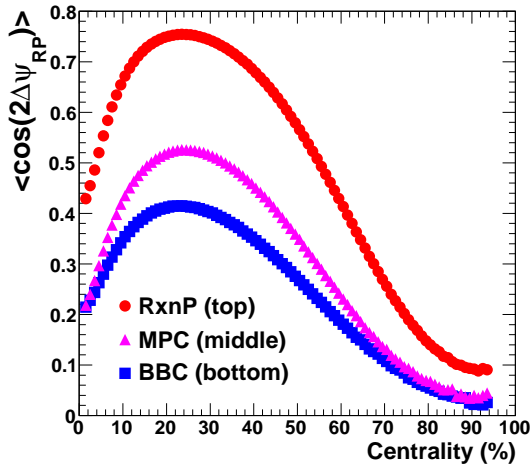}
}
\caption{\label{fig:RxnP_reso}Reaction plane resolution correction ($\langle cos(2\Delta\psi_{RP})\rangle$) as a function of centrality, measured with the RxnP in circles, the Muon Piston Calorimeters (MPC) in triangles, or the BBC in squares. Larger values correspond to higher precision in the reaction plane determination.}      
\end{figure}

Elliptic flow is measured with respect to the collision reaction plane estimated using a subset of particles produced during the collision~\cite{PhysRevC.58.1671,PhysRevD.46.229}. Careful attention must be taken to choose the subset so that all correlations other than the elliptic flow itself are removed between this subset and the particle of interest. Until 2007 the collision reaction plane was measured in PHENIX by using Beam-Beam counters (BBC). In 2007 a new reaction plane detector (RxnP) was installed. It allows two times better precision on the reaction plane measurement, as shown in Figure~\ref{fig:RxnP_reso}. The resolution by which the reaction plane angle is measured is used as a correction to the measured $v_2$ to account for the finite precision of the reaction plane measurement.

At forward rapidity, $\jpsi$s are measured through their decay into di-muons with two spectrometers, one on each side of the interaction point. Each spectrometer is composed of muon identification (MuID) and tracking (MuTr) systems~\cite{Adcox:2003zm}. The MuID, made of Iarroci tubes and steel absorbers, allows the identification of the muons through their penetration depth. The MuTr is made of cathode strip chambers that measure the particles momentum through their bending in a magnetic field. The muon spectrometers have a rapidity coverage for single muons of $|\eta|\in [1.2-2.2]$ which overlaps with the RxnP acceptance ($|\eta|\in[1-2.8]$). To eliminate possible bias from radiated gluons accompanying $\jpsi$s at similar rapidity, only the RxnP detector opposite to the arm where the muons go is used in this analysis.

\section{Analysis Method}

The integrated luminosity used for the analysis for this result is 537~$\mu$b$^{-1}$ (611~$\mu$b$^{-1}$) at negative (positive) rapidity, which is almost four times more than in the previous publications~\cite{Adare:2006ns}. 
This analysis is performed on an online-filtered data sample which enabled results very soon after the data was taken. The online filtering was based on fast tracking in the MuID and has no effect on the $v_2$ derived here.

The elliptic flow is obtained by dividing $\jpsi$s into only two $\phi-\psi$ bins: ]0,$\pi$/4] and ]$\pi/4$,$\pi$/2]. $v_2$ is then calculated by comparing the number of $\jpsi$s that belong to the bin which contains the reaction plane, $N^{in}$, with those that are in the bin which does not contain the reaction plane $N^{out}$, following the formulae: 

\begin{equation}
v_2^{\rm meas}=\frac{\pi}{4}\cdot \frac{(\nin-\nout)}{(\nin+\nout)}
\end{equation}
with uncertainty,
\begin{equation}
\sigma_{v_2^{\rm meas}}=\frac{\pi/2}{(\nin+\nout)^2}\cdot \sqrt{(\nout\sigmain)^2 + (\nin\sigmaout)^2 }
\end{equation} 

where $\sigma^{in}$ is the error on $\nin$ and $\sigma^{out}$ is the error on $\nout$. Finally, $v_2^{\rm meas}$ is divided by the reaction plane resolution in order to get the true $\jpsi$ $v_2$ as described in~\cite{PhysRevD.46.229}.

\begin{figure}[h]
\centering\resizebox{0.45\textwidth}{!}{%
  \includegraphics{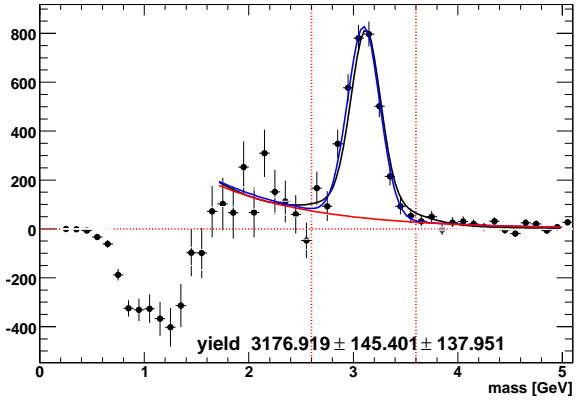}
}
\caption{\label{fig:mixed}20-60\% invariant mass distribution using the mixed-event background subtraction. The signal is obtained averaging results of three different fits~\cite{Adare:2006ns}.}
\end{figure}

Since the background subtraction is critical for the $v_2$ measurement, improvement on the subtraction method has been made for this analysis. Previously, a mixed-event subtraction technique was used to extract the $\jpsi$ signal in a high background environment~\cite{Adare:2006ns}. This method accounts for the fact that the MuTr acceptance differs for like-sign ($++ {\rm or} --$) and unlike-sign ($+-$) muon pairs and improves the statistical errors for bins where the signal over background is poor. The invariant mass distribution using the mixed-event subtraction for the 20-60\% centrality range is shown in Figure~\ref{fig:mixed}. The signal fit seems to be accurate, but the spectrum is distorted for masses below 2.6~GeV/c$^2$. The distortions originate from a bias introduced by the online filtering which creates a difference in the mixed-event ($Mixed$) and same-event ($FG$) distributions. To account for this bias, we use both like-sign and mixed-event distributions as follow:
\begin{equation}
FG_{+-}-FG_{++ {\rm or} --}\cdot \frac{Mixed_{+-}}{Mixed_{++ {\rm or} --}}
\end{equation}
The like-sign distributions help correcting for acceptance effects caused by the online filtering and the mixed-event distributions allow to reproduce a statistics unlimited background. 

\begin{figure}[h]
\centering\resizebox{0.45\textwidth}{!}{%
  \includegraphics{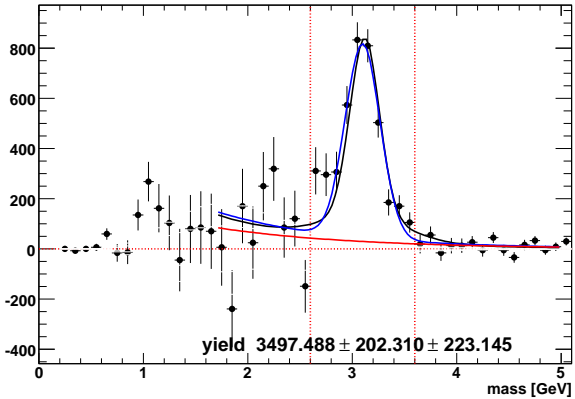}
}
\caption{\label{fig:combined}20-60\% invariant mass distribution using the combined background subtraction. The signal is obtained averaging results of three different fits~\cite{Adare:2006ns}.}
\end{figure}

The mass spectrum obtained with the combined subtraction is shown in Figure~\ref{fig:combined}. The low mass distortions are highly reduced and the residual background below the $\jpsi$ peak should also be reduced. On the other hand, this method results in statistical errors that are larger by a factor of about $\sqrt 2$ due to the use of the unlike-sign same-event distributions. 

\section{Results}

\begin{figure}[h]
\centering\resizebox{0.5\textwidth}{!}{%
  \includegraphics{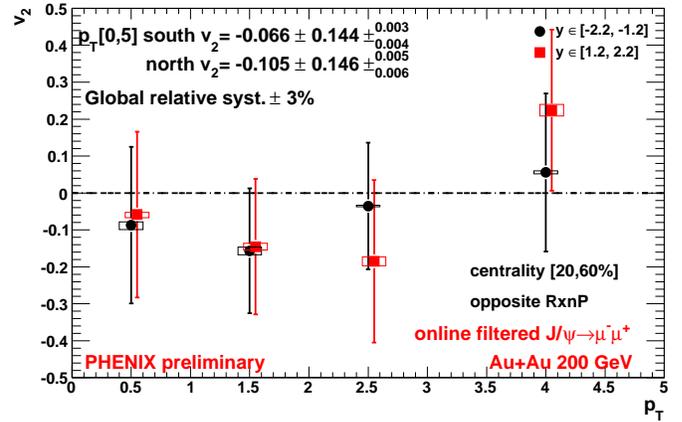}
}
\caption{\label{fig:2arms}$v_2$ as a function of $p_T$ for negative rapidity arm in black circles (red squares) at positive rapidity.}
\end{figure}

\begin{figure}[h]
\centering\resizebox{0.49\textwidth}{!}{%
  \includegraphics{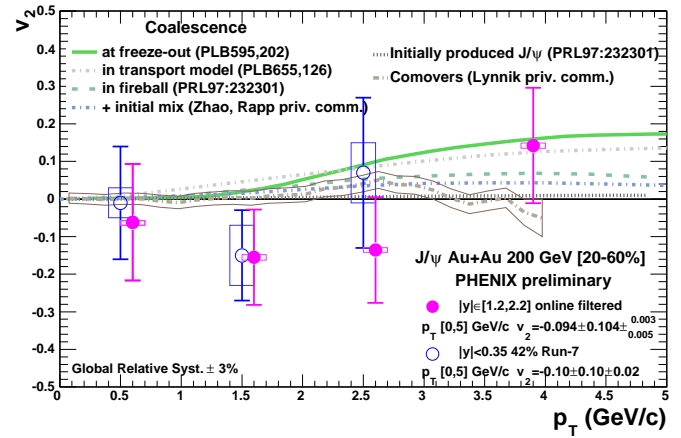}
}
\caption{\label{fig:results}$v_2$ as a function of $p_T$ for combined forward rapidity in magenta closed circles (full data sample) and at mid-rapidity in open blue circles (42\% of the data sample), with theoretical predictions made for mid-rapidity.}
\end{figure}

The elliptic flow at forward rapidity ($|y|\in[1.2-2.2]$) as a function of $p_T$ for the 20-60\% centrality selection is shown in Figure~\ref{fig:2arms} for negative (positive) rapidity in black circles (red squares). 
In Au+Au collisions, the forward-backward rapidity symmetry allows one to average the two measurements. The averaged result is shown by the magenta closed circles in Figure~\ref{fig:results}. The error bars on the two figures account for statistical and point-to-point uncorrelated errors. They come from the statistical uncertainty on the number of signal counts~\cite{Adare:2006ns}, and the systematic uncertainty of the signal counting and background line shapes. The boxes represent point-to-point correlated errors that account for the error on the average RxnP detector resolution and the error on the $\jpsi$ $\phi$ angle measurement (least accurate for $p_T<1$~GeV/c). An additional global relative uncertainty of 3\%, written on the figures, accounts for the error on the technique used to determine the reaction plane angle and resolution~\cite{PhysRevC.58.1671,PhysRevD.46.229}.

Also shown in Figure~\ref{fig:results} are PHENIX preliminary measurements at mid-rapidity ($|y|<0.35$)~\cite{Silvestre:2008tw} in open circles. Forward and mid-rapidity measurements are independent measurements since they use different detectors and triggers. In addition the methods to extract $v_2$ are also different and thus so are the uncertainties. At mid-rapidity, the point-to-point correlated systematic uncertainties are dominated by the unknown shape of the background $v_2$. At forward rapidity, only corrections for the finite precision of the reaction plane measurement and $\phi$ angle matter, after the total subtraction of the background. Forward and mid-rapidity measurements are perfectly compatible with each other for all $p_T$ bin. The $v_2$ obtained for each rapidity for $p_T\in[0-5]$~GeV/c is: $-0.094\pm0.104\pm^{0.003}_{0.005}$ at $|y|\in[1.2-2.2]$ and $-0.10\pm0.10\pm0.02$ at $|y|<0.35$. 

It is unclear whether $\jpsi$ elliptic flow at forward rapidity should differ from mid-rapidity, especially since the forward rapidity measurement only reaches $|y|=2.2$ and the underlying physics related to collective behavior might not differ much over such a rapidity range. Given this, an option is to combine the two rapidity measurements to increase statistical significance of the results. The statistical compatibility for $v_2$ to be positive for $p_T\in[0-5]$~GeV/c is only of the order of 10\%. It is dominated by the $[1-2]$~GeV/c bin which is the bin where the $\jpsi$s $p_T$ distribution is peaked. The probability of having a positive $v_2$ in $[1-2]$~GeV/c is only 6\%. It is more interesting to know if $v_2$ is positive and to what extent for $p_T>2$~GeV/c since coalescence models predict an increase of $\jpsi$ flow in this region as seen in Figure~\ref{fig:results}. The statistical probability for the combined $v_2$ measurements to be positive in $[2-5]$~GeV/c is 42\%, and for it to be above 0.1 is 15\%, which does not allow to distinguish between models. Additional data especially for $p_T>3$~GeV/c would give more statistical significance. 

Several predictions of $v_2$ are shown in Figure~\ref{fig:results}. Higher $v_2$ is predicted when more recombination is at play\cite{Zhu:2004nw,Greco:2003vf,Ravagli:2007xx,yan-2006-97}, and almost no $v_2$ for direct $\jpsi$s~\cite{yan-2006-97,Zhao:2007hh}, even when including absorption and comover interactions~\cite{Linnyk:2008uf}. It is to be noted that these predictions are for minimum bias collisions, and computed for mid-rapidity. No predictions at forward rapidity are yet available.

The $\jpsi$ $v_2$ measured by PHENIX can be but into perspective compared to recent lower energy measurements at the SPS. NA60 measured a positive $\jpsi$ $v_2$  of $7\pm3$\% in In+In non-central collisions~\cite{arnaldi} with no selection in $p_T$. At the SPS energy and in In+In collisions, given the much smaller charm cross section, it is unlikely that a sufficient number of $c \bar c$ pairs is produced to allow substantial formation of $\jpsi$ by recombination. If confirmed, the measured positive flow at the SPS is likely caused by other phenomena. One might speculate that nuclear absorption is responsible, despite earlier expectations that this effect should be negligible. Also at RHIC energies there may be additional contributions affecting v2. In any case, models should provide predictions for different nuclei and energies so that the measurements can distinguish any other effects from elliptic flow induced by recombination in a QGP. We note that NA50 measured a small $\jpsi$ $v_2$ in Pb+Pb collisions~\cite{prino} with a maximum of $3.5\pm1.5\pm1.3$\% for $p_T=2$~GeV/c. This result is not binned in centrality, and since $v_2$ for central collisions should be close to zero, their measurement could correspond to slightly larger values for non-central collisions. Interpretations of these results would also help to shed light on higher energy measurements done at RHIC.

\section{Outlook}

PHENIX has measured $\jpsi$ azimuthal anisotropy at forward rapidity for a centrality of 20-60\% using all the statistics from the 2007 data taking, and an online-filtered sample. Forward and mid-rapidity results are compatible within large uncertainties. For each measured $p_T$ bin, the value is compatible with both zero and maximal flow. The current precision on the measurements does not allow firm conclusions.

The forward rapidity statistical significance will be improved for final results by using the minimum bias rather than the online-filtered sample, with a gain in statistics by about $\sim$10\% and by benefiting from the improved handling of the combinatorial
background described earlier, leading to a reduction of the statistical uncertainty by $\sqrt 2$. A much larger data sample is expected to be available at RHIC in 2010.

\end{document}